\pdfoutput=1

\documentclass[11pt]{article}

\usepackage{ACL2023}

\usepackage{times}
\usepackage{latexsym}

\usepackage[T1]{fontenc}

\usepackage[utf8]{inputenc}

\usepackage{microtype}

\usepackage{inconsolata}
\usepackage[normalem]{ulem}
\usepackage{amsmath}
\usepackage{amssymb}
\usepackage{amstext}
\usepackage{bm}
\usepackage{bbm}
\usepackage{algorithm}
\usepackage{algpseudocode}
\usepackage{amsmath}
\usepackage{multirow}
\usepackage{graphicx}
\usepackage{subfigure}
\usepackage{epsfig}
\usepackage{float}
\usepackage{enumitem}
\usepackage{epstopdf}
\usepackage{framed}
\usepackage{url}
\usepackage{stfloats}
\usepackage{xspace}
\usepackage{booktabs}
\usepackage{subeqnarray}
\usepackage{enumerate}
\usepackage{color}

\newcommand{\paratitle}[1]{\vspace{1.5ex}\noindent\textbf{#1}}
\newcommand{\ie}{\emph{i.e.,}\xspace}
\newcommand{\aka}{\emph{a.k.a.,}\xspace}
\newcommand{\eg}{\emph{e.g.,}\xspace}

\newcommand{\ignore}[1]{}

\usepackage{makecell}

\usepackage{cite}

\title{TOME: A Two-stage Approach for Model-based Retrieval}

\author{\textbf{Ruiyang Ren\textsuperscript{1,3}\thanks{~~The work was done during the internship at Baidu.} \quad 
Wayne Xin Zhao\textsuperscript{1,3}\thanks{\llap{}\:\:\:Corresponding authors. } \quad 
Jing Liu\textsuperscript{2}\footnotemark[2] \quad  
Hua Wu\textsuperscript{2}}\\
\textbf{Ji-Rong Wen\textsuperscript{1,3} \quad  Haifeng Wang\textsuperscript{2}  
}\\
	\textsuperscript{1}Gaoling School of Artificial Intelligence, Renmin University of China  \\
	\textsuperscript{2}Baidu Inc. \\
	\textsuperscript{3}Beijing Key Laboratory of Big Data Management and Analysis Methods\\
	\{reyon.ren, jrwen\}@ruc.edu.cn, batmanfly@gmail.com\\
	\{liujing46, wu\_hua, wanghaifeng\}@baidu.com
}

\begin{document}
\maketitle
\sloppy
\begin{abstract}
Recently, model-based retrieval has emerged as a new paradigm in text retrieval that discards the index in the traditional retrieval model and instead memorizes the candidate corpora using model parameters. 
This design employs a sequence-to-sequence paradigm to generate document identifiers, which enables the complete capture of the relevance between queries and documents and simplifies the classic index-retrieval-rerank pipeline. 
Despite its attractive qualities, there remain several major challenges in model-based retrieval, including the discrepancy between pre-training and fine-tuning, and the discrepancy between training and inference.
To deal with the above challenges, we propose a novel two-stage model-based retrieval approach called TOME, which makes two major technical contributions, including the utilization of tokenized URLs as identifiers and the design of a two-stage generation architecture. 
We also propose a number of training strategies to deal with the training difficulty as the corpus size increases. 
Extensive experiments and analysis on MS MARCO and Natural Questions demonstrate the effectiveness of our proposed approach, and we investigate the scaling laws of TOME by examining various influencing factors. 

\end{abstract}

\section{Introduction}
Information retrieval systems have undergone continuous development over the past few decades, with the aim of obtaining relevant resources, such as documents, in response to a user query from a vast collection. 
With the recent success of Pre-trained Language Models (PLMs)~\citep{bert2019naacl,Raffel2020ExploringTL, zhao2023survey}, researchers have developed PLM-based dense retrievers~\citep{lin2021pretrained, DRSurvey}, 
which utilize dual-encoders and nearest neighbor search index for retrieval and achieve significant improvements over sparse retrievers. 

More recently, a new retrieval paradigm, known as \emph{model-based retrieval}~\citep{Tay2022TransformerMA, Zhou2022DynamicRetrieverAP}, has been introduced by developing an alternative architecture for retrieval. 
In contrast to traditional retrieval methods, it does not explicitly maintain a corpus index, thereby simplifying the classic \emph{index-retrieve-rerank} process. 
Typically, a model-based retrieval system is built based on a sequence-to-sequence generation model with an encoder-decoder architecture, such as T5~\citep{Raffel2020ExploringTL} and BART~\citep{Lewis2020BARTDS}.  
It accepts a query as input and directly generates the corresponding document identifier via the generation model.

Despite its attractive benefits in simplifying the retrieval pipeline, model-based retrieval still faces following major challenges. 
\begin{itemize}[leftmargin=1.2em]
\item Firstly, since the retrieval task is framed as a prediction task of document identifiers, making it crucial to design document identifiers that are well-suited to the underlying generative PLM. 
However, this issue is rarely discussed in prior research, and most existing approaches employ manually or randomly constructed identifiers~(\ie docids) as generation targets. 
Such docids are not adequately captured in the pre-training stage of the generative PLM, thus limiting PLM's capabilities for generative prediction (e.g., unseen docids during pre-training). 
This creates a discrepancy between the pre-training and fine-tuning phases. 
\item Secondly, there is a discrepancy between training and inference in the single-model generative architecture. 
While most existing studies incorporate multi-task learning~\citep{Tay2022TransformerMA} and auxiliary pre-training tasks~\citep{Ultron} to model both documents and queries during training, the model only processes queries during inference, resulting in a gap between the training and inference stages. 
\end{itemize}

\begin{figure*}
	\centering 
	\includegraphics[width=0.98\textwidth]{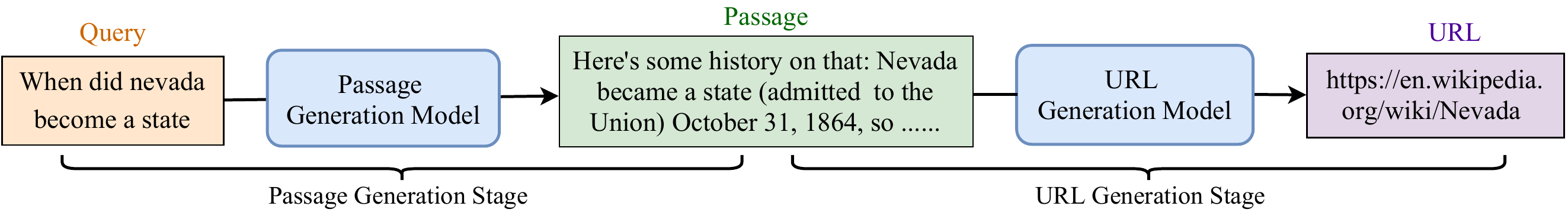}
	\caption{The illustration of the proposed two-stage generation approach. }
	\label{fig:model} 
\end{figure*}

To this end, in this paper, we propose a novel \underline{T}w\underline{O}-stage \underline{M}odel-based r\underline{E}trieval approach, \textbf{TOME} (as illustrated in Figure~\ref{fig:model}), which makes two major technical contributions. 
\begin{itemize}[leftmargin=1.2em]
    \item Firstly, we suggest using tokenized URLs (or URIs) as text identifiers, which are widely available for web pages or Wikipedia pages~\footnote{Regarding to other types of documents, we can use tokenized URIs as the identifiers.}. By using URL-based identifiers, the tokenized symbols are well aligned with the vocabulary of the generative PLM, thereby enhancing the generative capacity of the PLM. 
    URLs are typically comprised of normal text, as opposed to manually or randomly constructed identifiers. 
    As a result, such an identifier design can be used to help alleviate the gap between pre-training and fine-tuning. 
    \item Secondly, our approach decomposes the prediction task into two consecutive stages, namely passage generation and URL generation, which are fulfilled by two separate T5-based generation models, respectively. 
    The first stage aims to generate a relevant passage in the corpus based on the query, while the second stage aims to generate the corresponding URL of the generated passage from the first stage. 
    This two-stage architecture can reduce the discrepancy between training and inference. 
    In addition, the entire generation process is progressive. Consequently, the second stage is capable of tolerating errors that may be introduced by the preceding stage and generates correct URLs. 
\end{itemize}

Moreover, we discover that optimizing model-based retrieval becomes a challenging task when dealing with a vast corpus. As a result, we propose a number of improved training strategies to optimize the generation models, including query augmentation, passage length reduction, and model scaling.

To verify the effectiveness of TOME, we conduct extensive experiments on the publicly available MS MARCO and NQ datasets. 
Experimental results demonstrate the effectiveness of the proposed method, including the URL identifier design and the two-stage generation process. 
Additionally, case studies indicate that the second stage can tolerate errors induced by the first stage. 
Furthermore, we investigate the scaling laws of TOME by examining different model sizes, corpus sizes, and text lengths. 
We anticipate that these experimental results will facilitate further research on model-based retrieval.

\ignore{
The contributions of this paper can be summarized as follows:
\begin{itemize}[topsep=2pt,parsep=0pt,partopsep=0pt]
    \item Aiming at the issues existing in model-based retrieval, we propose corresponding improvement solutions, including using textual URLs as identifiers, two-stage generation process and optimization for data scaling problems.
    \item Experimental results on MS MARCO and NQ show that the proposed method significantly outperforms methods that using string docids as identifiers, and is competitive with dense retrieval methods.
    \item The in-depth experimental analyses provides a careful discussion of the proposed method and lead to some interesting conclusions.
\end{itemize}
}
\section{Related Works}
\paratitle{Text Retrieval.}
Text retrieval endeavors to find textual information related to a query from a large candidate corpus.
Early studies on sparse retrieval focused on term matching by utilizing sparse representations and inverted indices, such as BM25~\citep{robertson2009probabilistic}. In recent years, with the resurgence of neural networks and the emergence of pre-trained language models~(PLMs)~\citep{bert2019naacl, Raffel2020ExploringTL}, dense retrieval achieves better performance beyond traditional sparse retrieval on multiple tasks~\citep{colbert2020sigir, dpr2020, ANCE, rocketqa}. 
The dense retrieval and the technique of approximate nearest neighbor search have been widely adopted in various applications~\citep{oguz2020unified, pair, rocketqav2, asai2021xor, ren-zero, Zhou2022SimANS}. 

\paratitle{Model-based Retrieval.}
Both sparse retrieval and dense retrieval rely on explicit indices. Recently, 
researchers have proposed model-based retrieval~(\aka generative retrieval) models~\citep{metzler2021rethinking, Tay2022TransformerMA}. These methods consider model parameters as retrieval indices and directly generate the identifiers of related documents. Such an idea is initially proposed for entity retrieval~\citep{DeCao2021AutoregressiveER}, which autoregressively generates unique entity identifiers. Following this approach, researchers have introduced sequence-to-sequence encoder-decoder architecture for document retrieval~\citep{ Zhou2022DynamicRetrieverAP,Bevilacqua2022AutoregressiveSE,DSI-QG,NCI,GRLS,CorpusBrain, Ultron}. 
As discussed in the previous section, there still remain issues with model-based retrieval, including the discrepancy between pre-training and fine-tuning, and the discrepancy between training and inference. Our work tries to deal with these issues with a two-stage generation architecture with URL identifiers. 

\section{Approach}
In this section, we first introduce the task formulation, followed by the description of the proposed two-stage generation approach \textbf{TOME}.

\subsection{Task Formulation}

In this work, we consider the task of text retrieval, which aims to find relevant text resources  (\eg documents) related to a query from a large corpus. We further assume that these texts can be accessed by an associated URL\footnote{For the passages in a Web page, we can append specific postfix (\eg tab symbol) to the document URL or simply share the same document URL for within-document passages.} (or URI). 

To develop our approach, we adopt the recently proposed model-based paradigm for text retrieval~\citep{Tay2022TransformerMA, DSI-QG}.  
For retrieval, a model-based retrieval model takes a query $q$ as input and uses the text-to-text model to generate the identifier $y$ (length $n$) of the relevant document in an autoregressive manner, with the conditional probability:
\begin{equation}
    \label{eq:generate}
    \text{Pr}_\mathcal{M}(y|q)=\prod_{i=1}^n\text{Pr}_\mathcal{M}(y_i|y_{<i}, q),
\end{equation}
where $y_i$ denotes the $i$-th output token in the identifier $y$, $y_{<i}$ denotes the previous tokens $y_1, \ldots, y_{i-1}$, and $\mathcal{M}$ represents the PLM.
The identifier can be an atomic token or a string~\citep{Tay2022TransformerMA}. In our setting, it is assigned to an associated URL of a text (refer to Section~\ref{sec:url-def}).  
Typically, a generative pre-trained language model~(PLM) with an encoder-decoder architecture is employed to implement the text-to-text model (\eg T5), which is typically optimized by a cross-entropy loss as follows: 
\begin{eqnarray}
    \label{eq:loss}
	\mathcal{L}(\mathcal{M}) &=& -\log  \text{Pr}_\mathcal{M}(y|q) \nonumber \\ 
    &=& -\sum_{i=1}^n \log \big(\text{Pr}_\mathcal{M}(y_i|y_{<i}, q) \big). 
\end{eqnarray}

The key to model-based retrieval is to design a generative architecture that employs suitable document identifiers, and to develop effective training methods that can effectively associate queries to the identifiers of documents. 
Next, we expound our approach in detail.

\subsection{Model Architecture}
In this section, we first introduce the design of document identifiers, and then present the  two-stage generation architecture.

\subsubsection{Identifier Design}
\label{sec:url-def}
Existing studies typically use \emph{docids} to represent a document~\citep{Tay2022TransformerMA, DSI-QG}.  These docids are often randomly generated or manually constructed, which may not exist in real-world text corpora. However, the generative PLM is pre-trained based on large-scale text corpora, leading to a discrepancy between pre-training and fine-tuning.

Different from previous approaches, we consider a tokenized form of URLs as the docids. 
We directly treat the URL as a text string and tokenize it into a sequence of tokens using a T5 tokenizer. 
For instance, a sample URL  `https://en.wikipedia.org/wiki/Nevada' can be tokenized to \{`https',
`://',
`en',
`.',
`wikipedia',
`.',
`org',
`/',
`wiki',
`/',
`N',
`e',
`vada'\}.
We use the token sequence as the prediction target of the generative PLM, following the generation formula of Equation~\eqref{eq:generate}. 
It is worth noting that Ultron~\citep{Ultron} also uses URLs as identifiers, where a URL is reversed and only used as part of an identifier (also involving titles and domains).
As a comparison, we solely utilize tokenized URLs as the identifier, without any additional processing.

Compared to non-linguistic docids, URLs typically contain more meaningful tokens in the form normal text and widely exist in real-world text corpora, making them more suitable to modeling and prediction using generative PLMs. 
During decoding, we can directly adopt the general text decoding method to generate the URL, without resorting to limited search strategies such as constrained beam search~\citep{Tay2022TransformerMA, Bevilacqua2022AutoregressiveSE}.
Since these tokenized symbols often overlap among different URLs (\eg web pages from the same domains), they naturally derives semantic strings as the clustering method in DSI~\citep{Tay2022TransformerMA}.

\subsubsection{Two-stage Generation Architecture}

The objective of the generative model for retrieval is to establish a correlation between a query and its corresponding docid (\ie URL).
However, owing to the scarcity of annotated data, various improved strategies 
such as multi-task learning~\citep{Tay2022TransformerMA} or  pre-training~\citep{Ultron} have been proposed. Typically, a model processes both \emph{documents} and \emph{queries} during training, while it processes only \emph{queries} during inference, resulting in the discrepancy between training and inference. 
To tackle this issue, we propose a two-stage generation approach with two different generation models: one for passage generation and the other for  URL generation, as shown in Figure~\ref{fig:model}.

\paratitle{Passage Generation.}
In the first stage, we employ a T5-based passage generation model to map an input query to the passage content according to Equation~\eqref{eq:generate}.
The generated passage is anticipated as a relevant passage in the corpus that can provide an answer to the query.  
The objective of the passage generation model is to memorize the passages in the corpus, so as to generate the passages with utmost precision. 
It is trained with query-passage pairs, where each pair comprises a query and a passage from the document, along with the corresponding labeled URL. 
Different from existing methods~\citep{Tay2022TransformerMA, Bevilacqua2022AutoregressiveSE}, we do not utilize any data structure to restrict the decoding process and simply use greedy search to generate an individual result for a query in an autoregressive manner, which has a high decoding efficiency.
By incorporating the intermediate passage generation, our approach can mitigate the training-inference discrepancy that the query encoder also needs to process documents~\citep{Tay2022TransformerMA}.

\paratitle{URL Generation.} 
In the second stage, another T5-based PLM is employed to predict the corresponding URL as the retrieval result, utilizing the passage generated by the passage generation model as input. The URL is generated by means of greedy search decoding in a similar manner as in Equation~\eqref{eq:generate}.  
The URL generation model is trained with passage-URL pairs, where each pair comprises a passage and its corresponding URL.
The objective of the URL generation model is to memorize all the URLs in the corpus, so as to map a generated passage related to a query to a corresponding URL. 
Meanwhile, even if the generated passages contain some  irrelevant content or noise, this stage can still make reliable predictions since it can employ long passages as the context, rather than short queries.

Overall, such a two-stage generation approach can more effectively capture the semantic relatedness between queries and identifiers by both reducing the training-inference discrepancy and enriching the generation context, which is specifically tailored for model-based retrieval.

\subsection{Training}
For both the passage generation model and the URL generation model, we optimize them independently by utilizing the cross-entropy loss for optimizing standard T5 models, as shown in Equation~\eqref{eq:loss}. Nevertheless, optimizing model-based retrieval approaches~\citep{DSI-QG, NCI} is a challenging task as they essentially require memorizing the corpus information, and generating long text also poses challenges in model convergence.  In this part, we further propose several strategies for improving the training of our approach.

\paratitle{Query Augmentation}.  Generating pseudo queries is proven to be effective in improving the performance of model-based retrieval~\citep{NCI, DSI-QG}.
Here, we utilize query generation for constructing the training data for passage generation. 
Specifically, we take the passage collection as the corpus, and use an existing query generation model (\ie DocT5query~\citep{doctttttquery}) trained on the labeled dataset to generate multiple pseudo queries for each passage in the corpus. 
Following DSI-QG~\citep{DSI-QG}, we use the top-$k$ sampling strategy for query generation, and set $k$ up to 20.
The generated pseudo queries and their corresponding passages are then used to construct query-passage pairs as the training data for the passage generation model.
Such a query augmentation method can significantly increase the availability of training data, and also enhance the generalization capability of the model for different queries.

\paratitle{Reducing the Passage Length.}  
Since passages are much longer than URLs, passage generation is more complicated than URL generation. In the generation task, a more extensive generation target results in larger search space, which typically leads to a decrease in efficiency and effectiveness. 
While, in our approach, passage generation serves as an indirect step for predicting the URL, so that we consider reducing the passage length for improving the training efficiency. For this purpose, we shorten the maximum truncation length of the passage, from 128 to 32. However, reducing the passage length will probably results in a information loss, thus hurting the  generation performance. As the solution, we concatenate the title (a short text) and the shortened passage for enhancing the contained semantics.
We also add prompts before titles and passage contents like ``title:'' or ``passage:'' for better generation performance.

\paratitle{Increasing Model Scale.} Model-based retrieval requires a strong memorization capacity from the generative PLM, especially for our approach that involves a passage generation stage. Besides, scaling up the text corpus will significantly increase the difficulty of corpus memorization, 
and the PLM with a small parameter scale will have a limited memorization capacity when the data scale reaches a certain level.   
Considering the two aspects, we scale the model size accordingly and employ a larger PLM when necessary. Specifically, we use T5-large (the first stage is more difficult) and T5-base for the two stages of our approach on a small corpus (\eg subsets of MS MARCO), respectively. 
Further, we increase them to T5-3B and T5-large accordingly on a large corpus  (\eg the full set of MS MARCO). 
Besides the improved capacity, we find that using a larger model size is also useful  in improving the convergence rate (as detailed in Section~\ref{sec:scaling}). 

\section{Experimental Settings}
This section describes the major experimental settings, including datasets, evaluation metrics, baselines and implementation details.

\subsection{Datasets and Evaluation Metrics}

\paratitle{Datasets.} We conduct experiments on two public available datasets, namely \textit{MS MARCO} \citep{msmarco} Passage Ranking and \textit{Natural Questions} (NQ)~\citep{nq}. 
(1) \textit{MS MARCO} contains Bing search queries as well as passages from web documents, making it one of the largest web search datasets to date, with a full corpus of over 8.8 million passages. 
In addition, we also consider two subsets, each containing 100K and 1M passages, by following ~\citep{Tay2022TransformerMA, DSI-QG}. 
Based on the MS MARCO Question Answering dataset, we extract the URLs associated with the passages, selecting a random URL if a passage contains multiple URLs
(2) The \textit{NQ} dataset is a question answering dataset where the query data is collected from Google search logs, and the document data is from Wikipedia.
We use the NQ320K version by following NCI~\citep{NCI}, which contains 320K labeled query-document pairs and 100K documents. We collect abstracts of documents as intermediate-generated passages. 

\paratitle{Evaluation Metric.} Following previous works, we adopt {Hits@1} as the evaluation metric. 
This metric is calculated as the percentage of queries to which the top-1 generation result is positive. 
Since the outputs of models at different stages are either passage texts or URL texts, unlike the conventional MS MARCO evaluation by determining whether the retrieved identifiers are in the identifier label list, we evaluate the results by determining whether it is an exact match to the label text.

\subsection{Baselines}
For comparison, we chose the following baselines including sparse retrieval, dense retrieval, and model-based retrieval.

{BM25}~\citep{robertson2009probabilistic} is a classical \textit{sparse retriever} that uses the inverted index to find relevant passages by term overlap. 
{DPR}~\citep{dpr2020} and ANCE~\citep{ANCE} are two representative \textit{dense retrievers} that adopts dual-encoder architecture.
For \textit{model-based retrievers},
{DSI}~\citep{Tay2022TransformerMA} is a pioneer work for model-based retrieval that uses a sequence-to-sequence model to map the input query to the relevant docid. 
We use the open-source code released by DSI-QG for reproducing DSI baseline on MS MARCO.
SEAL~\citep{Bevilacqua2022AutoregressiveSE} is proposed to generate multiple ngrams for a query with an auxiliary Ferragina Manzini index.
{DSI-QG}~\citep{DSI-QG} proposes to improve DSI with augmented data 
constructed by query generation.
NCI~\citep{NCI} also utilizes pseudo queries for improving model-based retrieval with tailored architecture.
Due to the different experimental settings of different methods, we copy the performance values for some baselines on NQ in NCI and reproduce all of the baselines on MS MARCO under the same evaluation strategy. 
All the model-based retrieval baselines adopt the ``large'' version of PLMs. 

\subsection{Implementation Details}
We conduct our experiments with the deep learning framework PaddlePaddle~\citep{ma2019paddlepaddle} and natural language processing toolkit PaddleNLP~\citep{paddlenlp} on up to 32 NVIDIA Tesla A100 GPUs (with up to 80G RAM).

\paratitle{PLM.}
The generation models adopted in our work are initialized with different parameter scales of T5~\citep{Raffel2020ExploringTL}.
In the passage generation model, we use T5-3B for initialization on MS MARCO Full, and other models are initialized with T5-large.
In the URL generation model, we use T5-large for initialization on MS MARCO Full, and other models are initialized with T5-base.

\paratitle{Hyper-parameters.}
We adopt Adam optimizer with a learning rate of 5e-5, and train the models for a maximum of 3M steps with bf16 mixed precision strategy. 
The batchsize is set up to 128, 384 and 80 for T5-base, T5-large and T5-3B, respectively.
The maximal length of queries, passages and URLs are set as 32, 32 and 80, respectively.
The warm-up step is set as 100K and 10K for passage and URL generation task, respectively.

\paratitle{Query Augmentation.} 
We adopt the existing docT5query-large~\citep{doctttttquery} model that trained on MS MARCO training set, and generate 20 and 15 queries per passage for MS MARCO and NQ, respectively.
For training data, we only use pseudo-labeled data constructed by query generation on MS MARCO, and use both pseudo-labeled data and labeled data on NQ.

\section{Experimental Results and Analysis}
In this section, we report the experimental results of our proposed approach and conduct comprehensive empirical analysis.

\begin{table}[t]
    \small
    \centering
    \begin{tabular}{lccc}
    \toprule
    {\textbf{Methods}} &\textbf{100K} & \textbf{1M} & \textbf{Full}\\
    \midrule
    BM25~\citep{Anserini} & 58.01 & 35.20  & 17.05  \\
    DPR~\citep{dpr2020}        & {71.84}  & \textbf{52.52}  &  \textbf{29.54} \\
    DSI~\citep{Tay2022TransformerMA} & 11.75 & - & -  \\
    DSI-QG~\citep{DSI-QG} & 65.64 & 40.43 & - \\
    \midrule
    {TOME (single-stage)} & 66.46 & 43.04 & 19.32 \\
    {TOME (two-stage)} & \textbf{71.93} & 47.19  & 22.03    \\
    \bottomrule
    \end{tabular}
    \caption{The Hits@1 results of different methods on variant corpus scales of MSMARCO. }
    \label{tab:marco}
\end{table}

\begin{table}[t]
    \small
    \centering
    \begin{tabular}{lc}
    \toprule
    {\textbf{Methods}}  &  \textbf{Hits@1} \\
    \midrule
    BM25~\citep{Anserini}  & 15.11  \\
    ANCE~\citep{ANCE} & 52.63 \\
    DSI~\citep{Tay2022TransformerMA}  & 35.60 \\
    SEAL~\citep{Bevilacqua2022AutoregressiveSE}  & 59.93 \\    
    NCI~\citep{NCI}  & {66.23} \\
    DSI-QG~\citep{DSI-QG}  & 61.34 \\
    \midrule
    {TOME (single-stage)}  & 64.93\\
    {TOME (two-stage)}  & \textbf{66.64}  \\
    \bottomrule
    \end{tabular}
    \caption{The results of different methods on NQ320K. }
    \label{tab:nq}
\end{table}

\begin{table}[t]
    \small
    \centering
    \begin{tabular}{lcc}
    \toprule
    \multirow{2}{0pt}[-0.15em]{\textbf{Variants}}  & \textbf{MS MARCO} & \textbf{NQ} \\
    &  \textbf{ 100K} & \textbf{ 320K} \\
    \midrule
    TOME (two-stage) & \textbf{71.93} & \textbf{66.64}  \\
    \midrule
    w/o prompt  & 71.49 & 65.60  \\
    w/ increased maxlen & 71.80 & 65.15 \\
    w/ reduced pseudo query & 69.23 & 64.73 \\
    \bottomrule
    \end{tabular}
    \caption{The Hits@1 results of different variants of TOME on MS MARCO 100K and NQ320K.}
    \label{tab:ablation}
\end{table}

\subsection{Main Results}
Table~\ref{tab:marco} and Table~\ref{tab:nq} report the overall results on MS MARCO and NQ320K. Based on the results, we have the following observations:

\paratitle{Comparison with Model-based Retrievers.}
We observe that TOME consistently outperforms model-based retrievers on three subsets of MS MARCO and NQ320K datasets, thereby demonstrating the effectiveness of the proposed method.  
Moreover, NCI is a competitive baseline on NQ320K, which uses tailored decoder architecture, preprocessed semantic docid, and regularization on top of DSI-QG, while our method is simply trained with the standard T5 configuration without any additional processing.
We also discover that DSI-QG is unable to effectively converge when trained on the MS MARCO Full. We speculate that random non-linguistic docids become a bottleneck as the corpus scales up, while the loss can normally converge when using normal text (\eg URL) as a generation target.

\paratitle{Effect of Two-stage Generation Architecture.}
By simply substituting the generation target of DSI-QG from random string docids to URLs (single-stage of our method), the performance has been improved (refer to DSI-QG and TOME single-stage in Table~\ref{tab:marco} and~\ref{tab:nq}), indicating that natural language identifiers are more suitable for model-based retrieval tasks than non-linguistic docids. 
Furthermore, if we employ the two-stage generation that includes an intermediate step to generate passages before generating URLs, the performance will be further improved (refer to TOME single-stage and TOME two-stage in Table~\ref{tab:marco} and~\ref{tab:nq}). Such observation demonstrates that integrating passage generation in the process of model-based retrieval leads to better performance.

\paratitle{Comparison with Dense Retrievers.}
By adopting a series of training strategies, we successfully train TOME on large-scale corpora. However, although TOME outperforms dense retrieval methods on MS MARCO 100K and NQ320K, there still remains a performance gap when compared to DPR on larger corpora such as MS MARCO 1M and Full. 
This indicates that our method still has gaps compared to advanced dense retrieval methods when the corpus scales up.
Since the model-based method necessitates complete memorization of the entire corpus, it inherently possesses a disadvantage in larger-scale corpora when compared to dense retrievers, which needs to be further explored. 

\begin{table*}[!t]
\tiny
\begin{tabular}{p{1.5cm}|p{4.4cm}|p{4.1cm}|p{4.4cm}}

\toprule
\textbf{Query} & \textbf{Label passages} & \textbf{Other evidence passage in the corpus} & \textbf{Passages generated by passage generation model} \\
\midrule
\multirow{1}{1.5cm}[-2em]{Fin de siecle meaning} 
& Fin de si cle [(fann duh see-ek-luh)] The end of the nineteenth century; the phrase is French for end of the century. Fin de si cle is particularly used to describe \textit{\uline{the period's self-conscious artistic movements and a sophisticated despair that became popular at}}  \ldots
& - 
& Fin de si cle [(fann duh see-ek-luh)] The end of the nineteenth century; the phrase is French for end of the century. Fin de si cle is particularly used to describe \textit{\uline{a period of national despair that ended in the French Alps, after the French Revolution.}}\\
\midrule
\multirow{1}{1.5cm}[-3.1em]{Is ginger good for sore throat} 
&  Foods that heal a sore throat! Ginger: Lots of it eaten as close to raw as possible. \textbf{\uline{Ginger is an analgesic (a pain-killer) that may alleviate the pain associated with a sore throat. It is also a good antibacterial and antifungal and can help fight the infection causing your sore throat.}} I love ginger and use it a lot whether I have a sore throat or not. I keep a grater handy and grate fresh ginger into soups \ldots
& 1. Ginger. Ginger, often used for treating the common cold, can also help with bronchitis. \textit{\uwave{The anti-inflammatory and immune-boosting properties present in ginger help relieve irritated, inflamed or swollen bronchial tubes.}} 1  Add half a teaspoon each of ground ginger, cinnamon and cloves in a cup of hot water.  \ldots 
& \textit{\uwave{Ginger is commonly used to soothe sore throats because of its anti-inflammatory and immune-boosting properties.}} \textbf{\uline{Ginger is an analgesic (a pain-killer) that may alleviate the pain associated with a sore throat. It is also a good antibacterial and antifungal and can help fight the infection causing your sore throat.}} \ldots
\\
\bottomrule
\end{tabular}
\caption{The comparison of the labeled passages and generated passages. The evidence passages are not manually labeled but contain relevant content. The \uline{\textit{italic words with underline}} represents the different parts of two passages, the \uwave{\textit{italic words with wavy underline}} and \uline{\textbf{bold words with underline}} in different passages represent the reference parts.}
\label{tab:case}
\end{table*}

\subsection{Ablation Study}
In this section, we conduct an ablation study to examine the effectiveness of strategies in TOME. We report the results on MS MARCO 100K and NQ320K. 
Here, we consider three variants based on TOME for comparison: 
(a) \underline{\emph{w/o prompt}} removes the prompts before titles and passages;
(b) \underline{\emph{w/ increased maxlen}} increases the maximum truncated length of passage from 32 to 128;
(c) \underline{\emph{w/ reduced pseudo query}} reduces the amount of pseudo query to 10 per passage.

Table~\ref{tab:ablation} presents the results for variants of TOME. We can observe the following findings: 
(a) The performance drops in \underline{\emph{w/o prompt}}, demonstrating that adding prompts for identifying the title and passage is helpful for generating better results. 
(b) The performance drops in \underline{\emph{w/ increased maxlen}}, demonstrating that due to various training strategies, shortening the maximum truncated passage length does not bring performance loss but reduces the difficulty of training.
(c) The performance drops in \underline{\emph{w/ reduced pseudo query}}, demonstrating the effectiveness of generating a large number of pseudo queries for data augmentation.

\subsection{Analysis on Two-stage Generation}
In this section, we investigate the generation results of the passage generation model quantitatively and qualitatively to showcase the superiority of the proposed two-stage generation approach.

\subsubsection{Quantitative Analysis}
We quantitatively analyze the generation results on MSMARCO dev set with the passage generation models trained on MS MARCO 100K. 

First, we are surprised to find that on the entire dev set, the proportion of generated passages are the passages exist in the corpus is about 95\%. 
In cases where the model failed to generate labels correctly, about 85\% of the generated passages still exist in the corpus. 
This result indicates that \textbf{the model is capable of memorizing the corpus precisely} and is able to generate a retrieval-like result. 
Moreover, previous studies of dense retrieval reveal that there are a lot of false negatives in MSMARCO~\citep{rocketqa}. 
We also observe that approximately 80\% of the generation results that are not labeled as positives but appear in the corpus are false negatives, showing that \textbf{model-based retrieval suffers from the same issue of false negatives as dense retrieval}. Despite this, the passage generation model actually has strong generation capability.

\subsubsection{Qualitative Analysis}
To explore the generative capabilities of TOME, we conduct a case study on MSMARCO 100K, utilizing a maximum truncation length of 128 for better illustration.

Table~\ref{tab:case} gives two sampled queries, along with their corresponding label passages, evidence passages (if available) and generated passages. 
With respect to the first query, the generated passage is not exactly the same as the labeled passage.
In comparison with the labeled positive passage, the second half of the generated passage is altered. 
Despite the alteration in the generation passage, the URL generation model is still able to accurately map it to the correct URL, indicating that \textbf{the URL generation model can tolerate changes introduced by the passage generation model}. 
In the second example, the model extracts relevant content from both the label passage and the evidence passage, and then combines the contents to create the generated passage. 
It is interesting to observe that \textbf{the passage generation model is capable of summarizing multiple passages}.

\begin{figure*}[t]
	\centering
	\subfigure[Corpus scales.]{\label{fig:data_scale}
		\centering
		\includegraphics[width=0.27\textwidth]{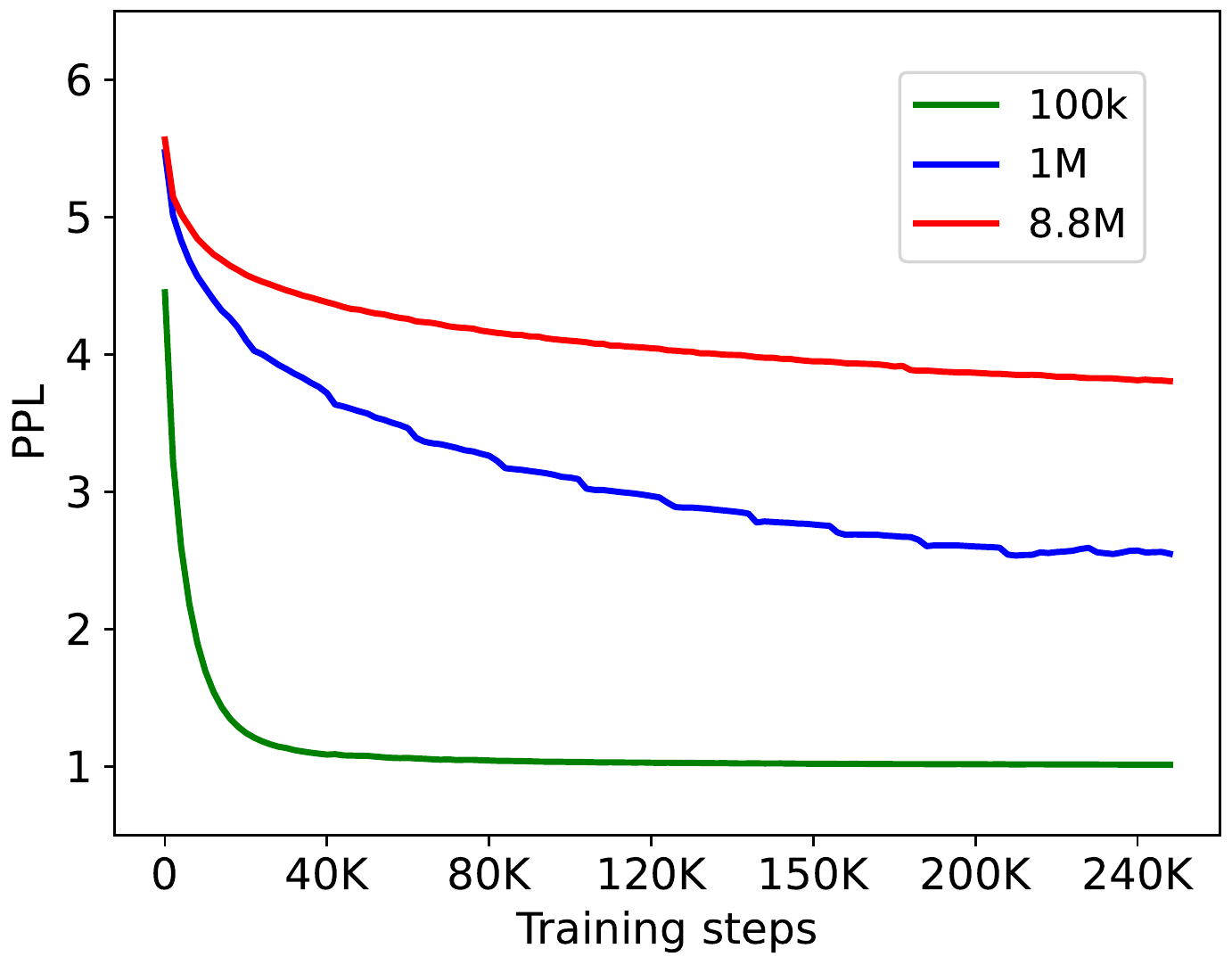}
	}\hspace{10mm}
	\subfigure[Model scales.]{\label{fig:model_scale}
		\centering
		\includegraphics[width=0.279\textwidth]{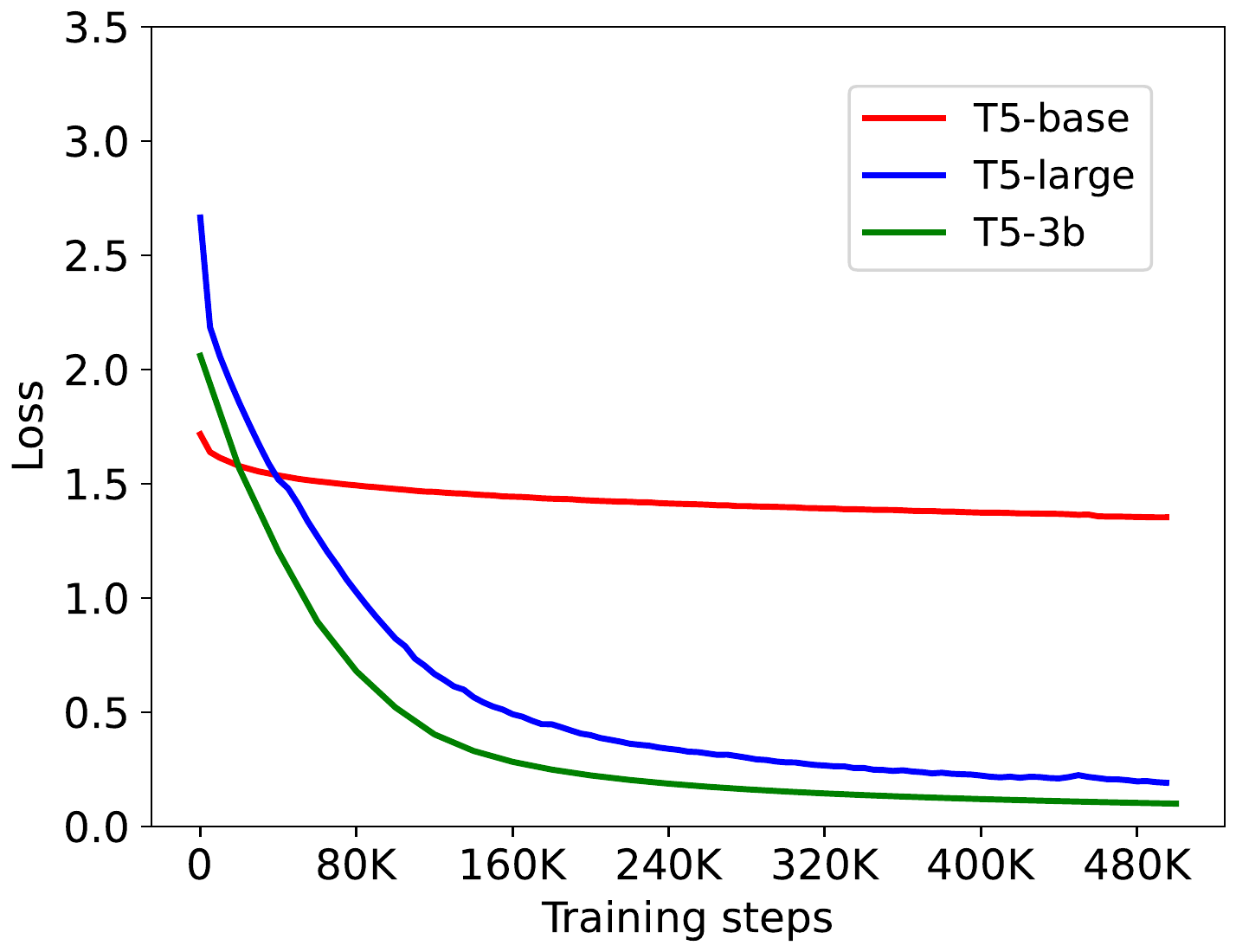}
	}\hspace{10mm}
	\subfigure[Maximum truncated length.]{\label{fig:joint-optimization}
		\centering
		\includegraphics[width=0.27\textwidth]{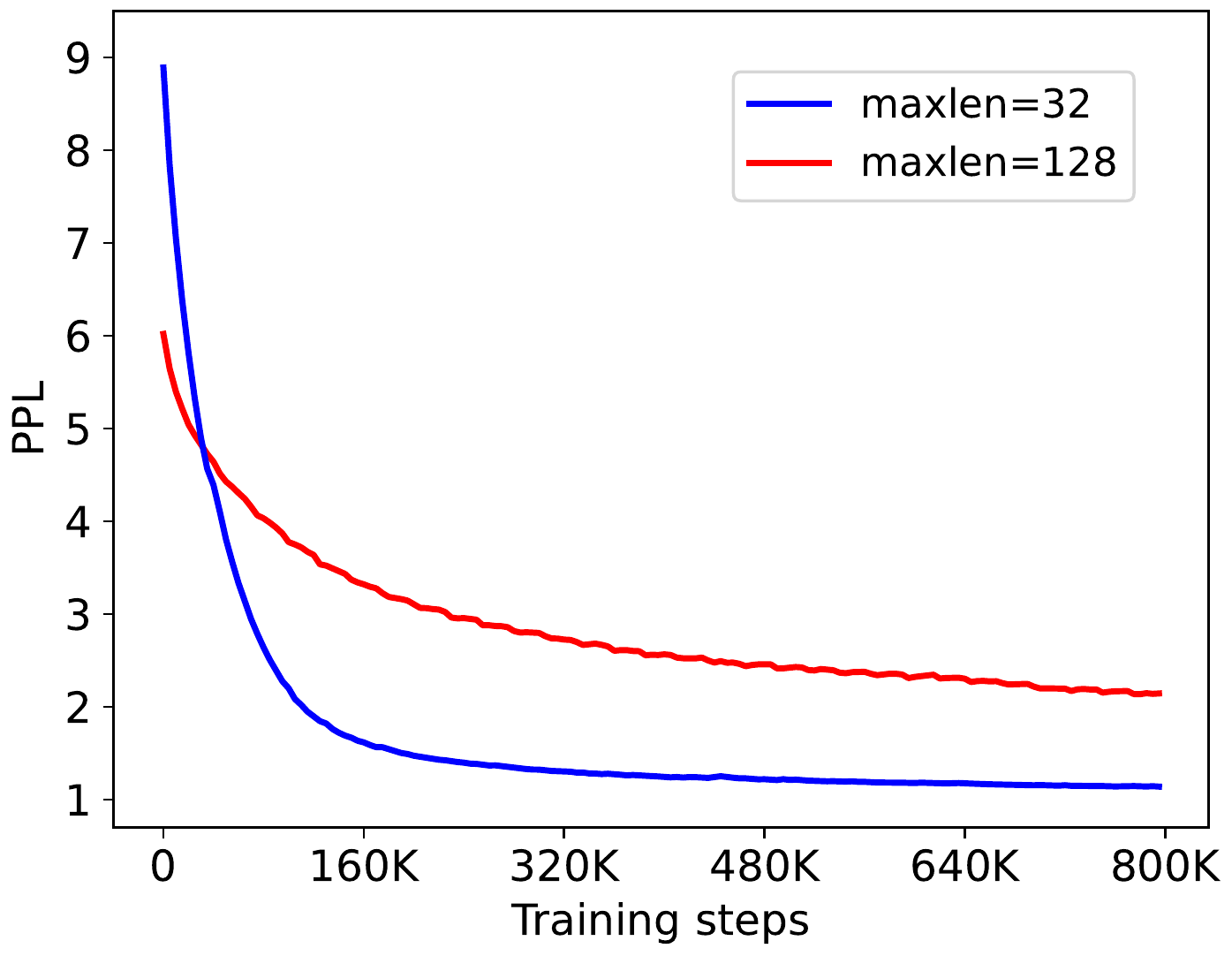}
	}
	\caption{The variation trends of model perplexity or training loss with the increase of training step under different corpus scales, model scales and maximum truncated length of passages.}
	\label{fig:para_len}
\end{figure*}

\subsection{Analysis on Scaling}
\label{sec:scaling}
We observe that long text generation poses a challenge to the convergence of loss, so we investigate the training efficiency and capability of the model under varying conditions. 
In particular, we use the same computing resource and conduct training on the passage generation stage (\ie the first stage) of TOME.
Considering that the trend is similar in the second stage, it has been omitted here due to limited space. 

\paratitle{Effect on Data Scale.} 
We investigate the impact of expanding the corpus on model training and examine whether the model capacity is insufficient when dealing with a large corpus. We fix the T5-large model and conduct training on MSMARCO 100K, 1M and Full datasets, respectively, without shortening the length of passages.
We use perplexity~(PPL) to estimate the model capacity and monitor how perplexity changes as training steps increase. 
The results are shown in Figure~\ref{fig:data_scale}.
It can be observed that the perplexity of the T5-large model fails to converge to a lower level after corpus scale expansion, which illustrates that under this task, a certain amount of data will lead to the capacity bottleneck of the model.
In addition, the decline rate of perplexity slows down on larger corpora, indicating that models with the same parameter size have low learning efficiency on a large-scale corpus..

\paratitle{Effect on Model Scale.}
To investigate the training characteristics of models with varying parameter scales, we fixed the data size to a intermediate scale of 1M, and used T5-base, T5-large and T5-3B models for training.
As depicted in Figure~\ref{fig:model_scale} shows the loss convergence of the model with different parameter sizes as training steps increases.
Among the three scales of models, the T5-base model exhibits the slowest convergence and encounters difficulties in convergence, while the T5-3B model with the largest scale of parameters converges the fastest. This indicates that PLMs with a larger number of parameters exhibit higher training efficiency, although with less data processing capability per step and more parameters to optimize under the same computing resources.

\paratitle{Effect on Passage Length.}
In order to investigate the effect of reducing the length of generated passages, we fixed the model as T5-large, and conducted experiments on passages with different maximum truncated lengths as generation targets on MSMARCO 1M.
Figure~\ref{fig:para_len} shows that after reducing the maximum truncated length of the generated passage, the perplexity significantly decreases, indicating that such a strategy is beneficial to mitigate the difficulty of the passage generation task.
Moreover, the model exhibited enhanced efficiency when generating shorter passages.

\section{Conclusion}

In this paper, we introduce TOME, a innovative two-stage model-based retrieval approach. To implement our approach, we make two major technical contributions in the design of the identifier and the architecture of two-stage generation. Moreover, we also employ a number of training strategies to better optimize our proposed architecture, especially on large-scale corpora. 
Extensive results demonstrate the effectiveness of TOME. Furthermore, we perform a thorough analysis and summarize the scaling law for the proposed method.
We believe such an idea itself is worthwhile for exploring in designing new model-based retrieval architecture.

\section*{Limitations}
In this work, we adopt a series of strategies for optimizing the generation models when corpus scaling up. Although we successfully train TOME on large-scale corpora, there is still a performance gap compared to mainstream dense retrieval methods under this scenario.
This is also one of the limitations of current model-based retrieval methods, because this retrieval paradigm requires the model to memorize the entire corpus, unlike dense retrievers that have strong generalization capability for different documents in a large corpus.
In addition, effective training on large-scale corpus also requires large-scale computing resources (up to 32 Tesla A100 80G GPU) and long training time, which will indirectly generate risks of energy consumption and emissions.


\bibliography{custom}

\begin{thebibliography}{33}
\expandafter\ifx\csname natexlab\endcsname\relax\def\natexlab#1{#1}\fi

\bibitem[{Asai et~al.(2021)Asai, Kasai, Clark, Lee, Choi, and
  Hajishirzi}]{asai2021xor}
Akari Asai, Jungo Kasai, Jonathan Clark, Kenton Lee, Eunsol Choi, and Hannaneh
  Hajishirzi. 2021.
\newblock {XOR} {QA}: Cross-lingual open-retrieval question answering.
\newblock In \emph{Proceedings of the 2021 Conference of the North American
  Chapter of the Association for Computational Linguistics: Human Language
  Technologies}, pages 547--564.

\bibitem[{Bevilacqua et~al.(2022)Bevilacqua, Ottaviano, Lewis, tau Yih, Riedel,
  and Petroni}]{Bevilacqua2022AutoregressiveSE}
Michele Bevilacqua, Giuseppe Ottaviano, Patrick Lewis, Wen tau Yih, Sebastian
  Riedel, and Fabio Petroni. 2022.
\newblock Autoregressive search engines: Generating substrings as document
  identifiers.
\newblock \emph{ArXiv}, abs/2204.10628.

\bibitem[{Cao et~al.(2021)Cao, Izacard, Riedel, and
  Petroni}]{DeCao2021AutoregressiveER}
Nicola~De Cao, Gautier Izacard, Sebastian Riedel, and Fabio Petroni. 2021.
\newblock Autoregressive entity retrieval.
\newblock In \emph{9th International Conference on Learning Representations,
  {ICLR} 2021, Virtual Event, Austria, May 3-7, 2021}.

\bibitem[{Chen et~al.(2022)Chen, Zhang, Guo, Liu, Fan, and Cheng}]{CorpusBrain}
Jiangui Chen, Ruqing Zhang, Jiafeng Guo, Yiqun Liu, Yixing Fan, and Xueqi
  Cheng. 2022.
\newblock \href {http://arxiv.org/abs/2208.07652} {Corpusbrain: Pre-train a
  generative retrieval model for knowledge-intensive language tasks}.
\newblock \emph{CoRR}, abs/2208.07652.

\bibitem[{Contributors(2021)}]{paddlenlp}
PaddleNLP Contributors. 2021.
\newblock Paddlenlp: An easy-to-use and high performance nlp library.
\newblock \url{https://github.com/PaddlePaddle/PaddleNLP}.

\bibitem[{Devlin et~al.(2019)Devlin, Chang, Lee, and Toutanova}]{bert2019naacl}
Jacob Devlin, Ming-Wei Chang, Kenton Lee, and Kristina Toutanova. 2019.
\newblock {BERT}: Pre-training of deep bidirectional transformers for language
  understanding.
\newblock In \emph{Proceedings of the 2019 Conference of the North {A}merican
  Chapter of the Association for Computational Linguistics: Human Language
  Technologies, Volume 1 (Long and Short Papers)}, pages 4171--4186.

\bibitem[{Karpukhin et~al.(2020)Karpukhin, Oguz, Min, Lewis, Wu, Edunov, Chen,
  and Yih}]{dpr2020}
Vladimir Karpukhin, Barlas Oguz, Sewon Min, Patrick Lewis, Ledell Wu, Sergey
  Edunov, Danqi Chen, and Wen-tau Yih. 2020.
\newblock Dense passage retrieval for open-domain question answering.
\newblock In \emph{Proceedings of the 2020 Conference on Empirical Methods in
  Natural Language Processing (EMNLP)}, pages 6769--6781.

\bibitem[{Khattab and Zaharia(2020)}]{colbert2020sigir}
Omar Khattab and Matei Zaharia. 2020.
\newblock Colbert: Efficient and effective passage search via contextualized
  late interaction over {BERT}.
\newblock In \emph{Proceedings of the 43rd International {ACM} {SIGIR}
  conference on research and development in Information Retrieval, {SIGIR}
  2020, Virtual Event, China, July 25-30, 2020}, pages 39--48.

\bibitem[{Kwiatkowski et~al.(2019)Kwiatkowski, Palomaki, Redfield, Collins,
  Parikh, Alberti, Epstein, Polosukhin, Devlin, Lee, Toutanova, Jones, Kelcey,
  Chang, Dai, Uszkoreit, Le, and Petrov}]{nq}
Tom Kwiatkowski, Jennimaria Palomaki, Olivia Redfield, Michael Collins, Ankur
  Parikh, Chris Alberti, Danielle Epstein, Illia Polosukhin, Jacob Devlin,
  Kenton Lee, Kristina Toutanova, Llion Jones, Matthew Kelcey, Ming-Wei Chang,
  Andrew~M. Dai, Jakob Uszkoreit, Quoc Le, and Slav Petrov. 2019.
\newblock Natural questions: A benchmark for question answering research.
\newblock \emph{Transactions of the Association for Computational Linguistics},
  7:452--466.

\bibitem[{Lee et~al.(2022)Lee, Yang, Oh, and Seo}]{GRLS}
Hyunji Lee, Sohee Yang, Hanseok Oh, and Minjoon Seo. 2022.
\newblock \href {http://arxiv.org/abs/2204.13596} {Generative retrieval for
  long sequences}.
\newblock \emph{CoRR}, abs/2204.13596.

\bibitem[{Lewis et~al.(2020)Lewis, Liu, Goyal, Ghazvininejad, Mohamed, Levy,
  Stoyanov, and Zettlemoyer}]{Lewis2020BARTDS}
Mike Lewis, Yinhan Liu, Naman Goyal, Marjan Ghazvininejad, Abdelrahman Mohamed,
  Omer Levy, Veselin Stoyanov, and Luke Zettlemoyer. 2020.
\newblock {BART}: Denoising sequence-to-sequence pre-training for natural
  language generation, translation, and comprehension.
\newblock In \emph{Proceedings of the 58th Annual Meeting of the Association
  for Computational Linguistics}, pages 7871--7880.

\bibitem[{Lin et~al.(2021)Lin, Nogueira, and Yates}]{lin2021pretrained}
Jimmy Lin, Rodrigo Nogueira, and Andrew Yates. 2021.
\newblock Pretrained transformers for text ranking: Bert and beyond.
\newblock \emph{Synthesis Lectures on Human Language Technologies},
  14(4):1--325.

\bibitem[{Ma et~al.(2019)Ma, Yu, Wu, and Wang}]{ma2019paddlepaddle}
Yanjun Ma, Dianhai Yu, Tian Wu, and Haifeng Wang. 2019.
\newblock Paddlepaddle: An open-source deep learning platform from industrial
  practice.
\newblock \emph{Frontiers of Data and Domputing}, 1(1):105--115.

\bibitem[{Metzler et~al.(2021)Metzler, Tay, Bahri, and
  Najork}]{metzler2021rethinking}
Donald Metzler, Yi~Tay, Dara Bahri, and Marc Najork. 2021.
\newblock Rethinking search: Making experts out of dilettantes.
\newblock \emph{CoRR}, abs/2105.02274.

\bibitem[{Nguyen et~al.(2016)Nguyen, Rosenberg, Song, Gao, Tiwary, Majumder,
  and Deng}]{msmarco}
Tri Nguyen, Mir Rosenberg, Xia Song, Jianfeng Gao, Saurabh Tiwary, Rangan
  Majumder, and Li~Deng. 2016.
\newblock {MS} {MARCO:} {A} human generated machine reading comprehension
  dataset.
\newblock In \emph{Proceedings of the Workshop on Cognitive Computation:
  Integrating neural and symbolic approaches 2016 co-located with the 30th
  Annual Conference on Neural Information Processing Systems {(NIPS} 2016),
  Barcelona, Spain, December 9, 2016}, volume 1773 of \emph{{CEUR} Workshop
  Proceedings}.

\bibitem[{Nogueira et~al.(2019)Nogueira, Lin, and Epistemic}]{doctttttquery}
Rodrigo Nogueira, Jimmy Lin, and AI~Epistemic. 2019.
\newblock From doc2query to doctttttquery.
\newblock \emph{Online preprint}.

\bibitem[{Oguz et~al.(2020)Oguz, Chen, Karpukhin, Peshterliev, Okhonko,
  Schlichtkrull, Gupta, Mehdad, and Yih}]{oguz2020unified}
Barlas Oguz, Xilun Chen, Vladimir Karpukhin, Stan Peshterliev, Dmytro Okhonko,
  Michael Schlichtkrull, Sonal Gupta, Yashar Mehdad, and Scott Yih. 2020.
\newblock Unik-qa: Unified representations of structured and unstructured
  knowledge for open-domain question answering.
\newblock \emph{arXiv preprint arXiv:2012.14610}.

\bibitem[{Qu et~al.(2021)Qu, Ding, Liu, Liu, Ren, Zhao, Dong, Wu, and
  Wang}]{rocketqa}
Yingqi Qu, Yuchen Ding, Jing Liu, Kai Liu, Ruiyang Ren, Wayne~Xin Zhao, Daxiang
  Dong, Hua Wu, and Haifeng Wang. 2021.
\newblock {R}ocket{QA}: An optimized training approach to dense passage
  retrieval for open-domain question answering.
\newblock In \emph{Proceedings of the 2021 Conference of the North American
  Chapter of the Association for Computational Linguistics: Human Language
  Technologies}, pages 5835--5847.

\bibitem[{Raffel et~al.(2020)Raffel, Shazeer, Roberts, Lee, Narang, Matena,
  Zhou, Li, and Liu}]{Raffel2020ExploringTL}
Colin Raffel, Noam~M. Shazeer, Adam Roberts, Katherine Lee, Sharan Narang,
  Michael Matena, Yanqi Zhou, Wei Li, and Peter~J. Liu. 2020.
\newblock Exploring the limits of transfer learning with a unified text-to-text
  transformer.
\newblock \emph{ArXiv}, abs/1910.10683.

\bibitem[{Ren et~al.(2021{\natexlab{a}})Ren, Lv, Qu, Liu, Zhao, She, Wu, Wang,
  and Wen}]{pair}
Ruiyang Ren, Shangwen Lv, Yingqi Qu, Jing Liu, Wayne~Xin Zhao, QiaoQiao She,
  Hua Wu, Haifeng Wang, and Ji-Rong Wen. 2021{\natexlab{a}}.
\newblock {PAIR}: Leveraging passage-centric similarity relation for improving
  dense passage retrieval.
\newblock In \emph{Findings of the Association for Computational Linguistics:
  ACL-IJCNLP 2021}, pages 2173--2183.

\bibitem[{Ren et~al.(2021{\natexlab{b}})Ren, Qu, Liu, Zhao, She, Wu, Wang, and
  Wen}]{rocketqav2}
Ruiyang Ren, Yingqi Qu, Jing Liu, Wayne~Xin Zhao, Qiaoqiao She, Hua Wu, Haifeng
  Wang, and Ji-Rong Wen. 2021{\natexlab{b}}.
\newblock Rocketqav2: A joint training method for dense passage retrieval and
  passage re-ranking.
\newblock In \emph{Proceedings of the 2021 Conference on Empirical Methods in
  Natural Language Processing}, pages 2825--2835.

\bibitem[{Ren et~al.(2022)Ren, Qu, Liu, Zhao, Wu, Ding, Wu, Wang, and
  Wen}]{ren-zero}
Ruiyang Ren, Yingqi Qu, Jing Liu, Wayne~Xin Zhao, Qifei Wu, Yuchen Ding, Hua
  Wu, Haifeng Wang, and Ji{-}Rong Wen. 2022.
\newblock \href {http://arxiv.org/abs/2204.12755} {A thorough examination on
  zero-shot dense retrieval}.
\newblock \emph{CoRR}, abs/2204.12755.

\bibitem[{Robertson et~al.(2009)Robertson, Zaragoza
  et~al.}]{robertson2009probabilistic}
Stephen Robertson, Hugo Zaragoza, et~al. 2009.
\newblock The probabilistic relevance framework: Bm25 and beyond.
\newblock \emph{Foundations and Trends{\textregistered} in Information
  Retrieval}, 3(4):333--389.

\bibitem[{Tay et~al.(2022)Tay, Tran, Dehghani, Ni, Bahri, Mehta, Qin, Hui,
  Zhao, Gupta, Schuster, Cohen, and Metzler}]{Tay2022TransformerMA}
Yi~Tay, Vinh~Quang Tran, Mostafa Dehghani, Jianmo Ni, Dara Bahri, Harsh Mehta,
  Zhen Qin, Kai Hui, Zhe Zhao, Jai Gupta, Tal Schuster, William~W. Cohen, and
  Donald Metzler. 2022.
\newblock Transformer memory as a differentiable search index.
\newblock \emph{ArXiv}, abs/2202.06991.

\bibitem[{Wang et~al.(2022)Wang, Hou, Wang, Miao, Wu, Sun, Chen, Xia, Chi,
  Zhao, Liu, Xie, Sun, Deng, Zhang, and Yang}]{NCI}
Yujing Wang, Yingyan Hou, Haonan Wang, Ziming Miao, Shibin Wu, Hao Sun,
  Qi~Chen, Yuqing Xia, Chengmin Chi, Guoshuai Zhao, Zheng Liu, Xing Xie,
  Hao~Allen Sun, Weiwei Deng, Qi~Zhang, and Mao Yang. 2022.
\newblock \href {http://arxiv.org/abs/2206.02743} {A neural corpus indexer for
  document retrieval}.
\newblock \emph{CoRR}, abs/2206.02743.

\bibitem[{Xiong et~al.(2021)Xiong, Xiong, Li, Tang, Liu, Bennett, Ahmed, and
  Overwijk}]{ANCE}
Lee Xiong, Chenyan Xiong, Ye~Li, Kwok{-}Fung Tang, Jialin Liu, Paul~N. Bennett,
  Junaid Ahmed, and Arnold Overwijk. 2021.
\newblock Approximate nearest neighbor negative contrastive learning for dense
  text retrieval.
\newblock In \emph{9th International Conference on Learning Representations,
  {ICLR} 2021, Virtual Event, Austria, May 3-7, 2021}.

\bibitem[{Yang et~al.(2017)Yang, Fang, and Lin}]{Anserini}
Peilin Yang, Hui Fang, and Jimmy Lin. 2017.
\newblock Anserini: Enabling the use of lucene for information retrieval
  research.
\newblock In \emph{Proceedings of the 40th International {ACM} {SIGIR}
  Conference on Research and Development in Information Retrieval, Shinjuku,
  Tokyo, Japan, August 7-11, 2017}, pages 1253--1256.

\bibitem[{Zhao et~al.(2022)Zhao, Liu, Ren, and Wen}]{DRSurvey}
Wayne~Xin Zhao, Jing Liu, Ruiyang Ren, and Ji-Rong Wen. 2022.
\newblock Dense text retrieval based on pretrained language models: A survey.
\newblock \emph{arXiv preprint arXiv:2211.14876}.

\bibitem[{Zhao et~al.(2023)Zhao, Zhou, Li, Tang, Wang, Hou, Min, Zhang, Zhang,
  Dong et~al.}]{zhao2023survey}
Wayne~Xin Zhao, Kun Zhou, Junyi Li, Tianyi Tang, Xiaolei Wang, Yupeng Hou,
  Yingqian Min, Beichen Zhang, Junjie Zhang, Zican Dong, et~al. 2023.
\newblock A survey of large language models.
\newblock \emph{arXiv preprint arXiv:2303.18223}.

\bibitem[{Zhou et~al.(2022{\natexlab{a}})Zhou, Gong, Liu, Zhao, Shen, Dong, Lu,
  Majumder, Wen, Duan, and Chen}]{Zhou2022SimANS}
Kun Zhou, Yeyun Gong, Xiao Liu, Wayne~Xin Zhao, Yelong Shen, Anlei Dong,
  Jingwen Lu, Rangan Majumder, Ji-Rong Wen, Nan Duan, and Weizhu Chen.
  2022{\natexlab{a}}.
\newblock Simans: Simple ambiguous negatives sampling for dense text retrieval.
\newblock In \emph{EMNLP}.

\bibitem[{Zhou et~al.(2022{\natexlab{b}})Zhou, Yao, Dou, Wu, Zhang, and
  Wen}]{Ultron}
Yujia Zhou, Jing Yao, Zhicheng Dou, Ledell Wu, Peitian Zhang, and Ji{-}Rong
  Wen. 2022{\natexlab{b}}.
\newblock \href {http://arxiv.org/abs/2208.09257} {Ultron: An ultimate
  retriever on corpus with a model-based indexer}.
\newblock \emph{CoRR}, abs/2208.09257.

\bibitem[{Zhou et~al.(2022{\natexlab{c}})Zhou, Yao, Dou, Wu, and rong
  Wen}]{Zhou2022DynamicRetrieverAP}
Yujia Zhou, Jing Yao, Zhicheng Dou, Ledell~Yu Wu, and Ji~rong Wen.
  2022{\natexlab{c}}.
\newblock Dynamicretriever: A pre-training model-based ir system with neither
  sparse nor dense index.
\newblock \emph{ArXiv}, abs/2203.00537.

\bibitem[{Zhuang et~al.(2022)Zhuang, Ren, Shou, Pei, Gong, Zuccon, and
  Jiang}]{DSI-QG}
Shengyao Zhuang, Houxing Ren, Linjun Shou, Jian Pei, Ming Gong, Guido Zuccon,
  and Daxin Jiang. 2022.
\newblock \href {http://arxiv.org/abs/2206.10128} {Bridging the gap between
  indexing and retrieval for differentiable search index with query
  generation}.
\newblock \emph{CoRR}, abs/2206.10128.

\end{thebibliography}
\bibliographystyle{acl_natbib}

\appendix



\end{document}